# A description of pseudorapidity distributions in *p-p* collisions at center-of-mass energy from 23.6 to 900 GeV[*]

JIANG Zhi-Jin (姜志进)[†], WANG Jie (王杰),

ZHANG Hai-Li (张海利) and MA Ke (马可)

College of Science, University of Shanghai for Science and Technology, Shanghai 200093, China

**Abstract** In the context of combined model of evolution-dominated hydrodynamics + leading particles, we discuss the pseudorapidity distributions of charged particles produced in *p-p* collisions. A comparison is made between the theoretical predictions and experimental measurements. The combined model works well in *p-p* collisions in the whole available energy region from $\sqrt{s}$ =23.6 to 900 GeV.

Ⅰ. **Introduction**

Owing to the success in characterizing the elliptic flow and multiplicity production in hadron or nucleus collisions [1-3], relativistic hydrodynamics has now been widely taken as one of the best tools for understanding the space-time evolution of the matter created in collisions [4-19].

Provided the initial conditions and the equation of state are given, the evolution of fluid relies only on the local energy-momentum conservation and the assumption of local thermal equilibrium. From this point of view, hydrodynamics is simple and powerful. However, on the other hand, the initial conditions and the equation of state are not well known. Worse still is that the partial differential equations of relativistic hydrodynamics are highly non-linear and coupled. It is a very hard thing to solve them analytically. From this point of view, hydrodynamics is tremendously complicated. That is why the most analytical work is, up till now, only limited to the hydrodynamics of 1+1 dimensions, which was first considered by Landau in the context of high energy collisions [20]. The 3+1 dimensional hydrodynamics is less developed, and no general exact solutions are known so far.

---

[*] Partly supported by the Transformation Project of Science and Technology of Shanghai Baoshan District with Grant No. CXY-2012-25, the Shanghai Leading Academic Discipline Project with Grant No. XTKX 2012 and the National Training Project with Grant No. 14XPM03.

[†] E-mail: Jzj265@163.com

One of the important applications of 1+1 dimensional hydrodynamics is the analysis of the pseudorapidity distributions of the charged particles produced in hadron or nucleus collisions. In our previous work [8], we have once successfully used the combined model of evolution-dominated hydrodynamics + leading particles in describing such distributions in nucleus-nucleus collisions at BNL-RHIC and CERN-LHC energies. Now, what we are concerned with is that whether the model can still work in hadron, such as in *p-p* collisions. To clarify this problem is just the subject of this paper. We can see that, just as in nucleus-nucleus collisions, the total contributions from both evolution-dominated hydrodynamics and leading particles are in good accordance with the experimental data measured in *p-p* collisions at available energies from $\sqrt{s}$ =23.6 to 900 GeV [21-24].

## 2. A brief description of the model

The 1+1 expansion of a perfect fluid obeys equation

$$\begin{cases} \dfrac{e^{2y}-1}{2}\dfrac{\partial(\varepsilon+p)}{\partial z^{+}}+e^{2y}(\varepsilon+p)\dfrac{\partial y}{\partial z^{+}}+\dfrac{1-e^{-2y}}{2}\dfrac{\partial(\varepsilon+p)}{\partial z^{-}} \\ \qquad\qquad\qquad +e^{-2y}(\varepsilon+p)\dfrac{\partial y}{\partial z^{-}}+\dfrac{\partial p}{\partial z^{+}}-\dfrac{\partial p}{\partial z^{-}}=0, \\ \dfrac{e^{2y}+1}{2}\dfrac{\partial(\varepsilon+p)}{\partial z^{+}}+e^{2y}(\varepsilon+p)\dfrac{\partial y}{\partial z^{+}}+\dfrac{1+e^{-2y}}{2}\dfrac{\partial(\varepsilon+p)}{\partial z^{-}} \\ \qquad\qquad\qquad -e^{-2y}(\varepsilon+p)\dfrac{\partial y}{\partial z^{-}}-\dfrac{\partial p}{\partial z^{+}}-\dfrac{\partial p}{\partial z^{-}}=0, \end{cases} \quad (1)$$

where $\varepsilon$, $p$ and $y$ are respectively the energy density, pressure and ordinary rapidity of fluid. $z^{\pm}=t\pm z=x^{0}\pm x^{1}=\tau e^{\pm\eta}$ is the light-cone coordinates, $\tau=\sqrt{z^{+}z^{-}}$ is the proper time, and $\eta=1/2\ln(z^{+}/z^{-})$ is the space-time rapidity of fluid.

Eq. (1) is a complicated, non-linear and coupled one. In order to solve it, one introduces Khalatnikov potential $\chi$, which relates to $z^{\pm}$, $\tau$ and $\eta$ by equations

$$\begin{cases} z^{\pm}(\theta,y)=\dfrac{1}{2T_{0}}e^{\theta\pm y}\left(\dfrac{\partial\chi}{\partial\theta}\pm\dfrac{\partial\chi}{\partial y}\right), \\ \tau(\theta,y)=\dfrac{e^{\theta}}{2T_{0}}\sqrt{\left(\dfrac{\partial\chi}{\partial\theta}\right)^{2}-\left(\dfrac{\partial\chi}{\partial y}\right)^{2}}, \\ \eta(\theta,y)=y+\dfrac{1}{2}\ln\left(\dfrac{\partial\chi/\partial\theta+\partial\chi/\partial y}{\partial\chi/\partial\theta-\partial\chi/\partial y}\right), \end{cases} \quad (2)$$

where

$$\theta=\ln\left(\dfrac{T_{0}}{T}\right), \quad (3)$$

$T$ is the temperature of liquid, and $T_0$ is its initial scale. In terms of $\chi$, Eq. (1) can be reduced to

$$\frac{\partial^2 \chi(\theta,y)}{\partial \theta^2} - [g(\theta)-1]\frac{\partial \chi(\theta,y)}{\partial \theta} - g(\theta)\frac{\partial^2 \chi(\theta,y)}{\partial y^2} = 0, \qquad (4)$$

where $1/\sqrt{g(\theta)} = c_s(\theta)$ is the speed of sound. The above equation is now a linear second-order partial differential equation, which works for any form of $g(\theta)$.

Investigations have shown that the speed of sound changes very slowly with interaction energy [25-27]. As an approximation, in the energy region we are concerned with in this paper, it can be well taken as a constant, that is $g(\theta) = g$. In this case, Eq. (4) has the solution as

$$\chi(\theta,y) = \frac{e^{\frac{g-1}{2}\theta}}{4\sqrt{g}}\int dy'\int_0^{\theta-(y-y')/\sqrt{g}} d\theta' F(\theta',y') I_0\left(\frac{g-1}{2}\sqrt{(\theta-\theta')^2 - \frac{(y-y')^2}{g}}\right), \qquad (5)$$

where $F(\theta',y')$ stands for the initial distribution of the sources of hydrodynamic flow. The specific value of $g$ can be fixed by fitting with experimental data.

In collisions at high energy, owing to the violent compression and Lorentz contraction of interaction system along beam direction, the initial pressure gradient of created matter in this direction is very large. By contrast, the effect of initial flow of sources is negligible. The motion of liquid is mainly dominated by the following evolution. In this evolution-dominated picture, the initial distribution of source takes the form as [4, 28, 29]

$$F(\theta',y') = Ce^{-\frac{g+1}{2}\theta'}\Theta(\theta')\delta(y'), \qquad (6)$$

where $C$ is a constant. Inserting it into Eq. (5), it reads

$$\chi(\theta,y) = Ce^{-\theta}\int_{y/\sqrt{g}}^{\theta} d\theta' e^{\frac{g+1}{2}\theta'} I_0\left(\frac{g-1}{2}\sqrt{\theta'^2 - \frac{y^2}{g}}\right). \qquad (7)$$

Along with the expansion of fluid, it will become cooler and cooler. As its temperature drops to a certain degree, the fluid will freeze out into the charged particles. If we assume that the freeze-out of fluid takes place at a space-like hypersurface with a fixed temperature of $T_{\text{FO}}$, and the number of charged particles is proportional to entropy, we can get the rapidity distribution of charged particles

$$\frac{dN_{\text{Fluid}}}{dy} \propto \left.\frac{\partial^2 \chi(\theta,y)}{\partial \theta^2} + \frac{\partial \chi(\theta,y)}{\partial \theta}\right|_{\theta=\theta_{\text{FO}}}, \qquad (8)$$

where $\theta_{\text{FO}} = \ln(T_0/T_{\text{FO}})$, which is related to the initial temperature of fluid and is therefore dependent on the incident energy. Its specific value can be determined by comparing with experimental data.

Substituting Eq. (7) into above equation, it becomes

$$\frac{dN_{\text{Fluid}}(\sqrt{s},y)}{dy} = C(\sqrt{s}) \left[ I_0\left(\frac{g-1}{2}\sqrt{\theta_{\text{FO}}^2 - \frac{y^2}{g}}\right) + \frac{\theta_{\text{FO}}}{\sqrt{\theta_{\text{FO}}^2 - y^2/g}} I_1\left(\frac{g-1}{2}\sqrt{\theta_{\text{FO}}^2 - \frac{y^2}{g}}\right) \right], \qquad (9)$$

where $C(\sqrt{s})$ is an overall normalization constant, $\sqrt{s}$ is the center-of-mass incident energy.

Apart from the charged particles resulted from the freeze-out of liquid, leading particles also have contributions to the produced charged particles [30-34]. In *p+p* collisions, there are only two leading particles. One is in projectile fragmentation region, and the other is in target fragmentation region. Considering that, for a given incident energy, the leading particles in each time of *p+p* collisions have approximately the same amount of energy, then, according to the central limit theorem [35], the leading particles should follow the Gaussian rapidity distribution. That is

$$\frac{dN_{\text{Lead}}(y,\sqrt{s})}{dy} = \frac{1}{\sqrt{2\pi}\sigma} \exp\left\{ -\frac{\left[|y| - y_0(\sqrt{s})\right]^2}{2\sigma^2} \right\}, \qquad (10)$$

where $\sigma$ and $y_0(\sqrt{s})$ are respectively the width and central position of Gaussian distribution. In fact, as is known to all, the rapidity distribution of any charged particles produced in hadron or heavy collisions can be well represented by Gaussian form [36-38]. Seeing that, for leading particles resulted in *p+p* collisions at different energies, the relative energy differences among them should not be too much, $\sigma$ should be independent of energy and therefore approximately remain a constant. It is evident that $y_0(\sqrt{s})$ should increase with energy. Both $\sigma$ and $y_0(\sqrt{s})$ can now only be determined by comparing with experimental data.

## 3. Comparison with experimental measurements

Having the rapidity distribution of Eqs. (9) and (10), the pseudorapidity distribution measured in experiments can be expressed as [39]

$$\frac{dN(\eta,\sqrt{s_{\text{NN}}})}{d\eta} = \sqrt{1 - \frac{m^2}{m_{\text{T}}^2 \cosh^2 y}} \frac{dN(y,\sqrt{s_{\text{NN}}})}{dy}, \qquad (11)$$

$$y = \frac{1}{2}\ln\left[\frac{\sqrt{p_{\text{T}}^2 \cosh^2 \eta + m^2} + p_{\text{T}} \sinh \eta}{\sqrt{p_{\text{T}}^2 \cosh^2 \eta + m^2} - p_{\text{T}} \sinh \eta}\right], \qquad (12)$$

where $p_{\text{T}}$ is the transverse momentum, $m_{\text{T}} = \sqrt{m^2 + p_{\text{T}}^2}$ is the transverse mass, and

$$\frac{dN(y,\sqrt{s_{NN}})}{dy} = \frac{dN_{Liquid}(y,\sqrt{s_{NN}})}{dy} + \frac{dN_{Lead}(y,\sqrt{s_{NN}})}{dy} \tag{13}$$

is the total rapidity distribution from both liquid evolution and leading particles.

Experiments have shown that the overwhelming majority of the produced charged particles in hadron or heavy ion collisions at high energy consists of pions, kaons and protons with proportions of about 83%, 12% and 5%, respectively [40]. Furthermore, the transverse momentum $p_T$ changes very slowly with beam energies [41, 42]. In calculations, the $p_T$ in Eqs. (11) and (12) takes the values *via* relation [41]

$$p_T = 0.413 - 0.0171\ln(s) + 0.00143\ln^2(s),$$

where $p_T$ and $\sqrt{s}$ are respectively in unit of GeV/c and GeV. The $m$ in Eqs. (11) and (12) takes the values from 0.20 to 0.28 GeV for energies from 23.6 to 900 GeV, which are approximately the mean masses of pions, kaons, and protons.

Substituting Eq. (13) into Eq. (11), we can get the pseudorapidity distributions of the produced charged particles. Figure 1 shows such distributions for *p-p* collisions at $\sqrt{s}$ = 23.6, 45.2, 200, 410, 546 and 900 GeV, respectively. The solid dots are the experimental measurements [21-24]. The dashed curves are the results got from the evolution-dominated hydrodynamics of Eq. (9). The dotted curves are the results obtained from leading particles of Eq. (10). The solid curves are the results achieved from Eq. (13), that is, the sums of dashed and dotted curves. The corresponding $\chi^2/\text{NDF}$ is 0.27, 0.31, 0.15 and 0.32 for $\sqrt{s}$ =23.6, 45.2, 200 and 900 GeV, respectively. It can be seen that the theoretical results are well consistent with experimental measurements.

In calculations, the parameter $g$ in Eq. (9) takes value of $g = 12$, or $c_s = 0.29$. $\theta_{FO}$ takes the values of 1.06, 1.36, 2.89, 3.81, 3.89 and 4.43 for energies from small to large. It can be seen that, $\theta_{FO}$ increases with energy. The central parameter $y_0(\sqrt{s})$ in Eq. (10) takes the values of 2.05, 2.14, 2.72, 2.83, 2.91 and 2.96 for energies from small to large. The $\sigma$ in Eq. (10) takes the value of 0.90 for all concerned incident energies. As the analyses given above, $y_0(\sqrt{s})$ increases with energy, but $\sigma$ remains a constant for different incident energies.

## 4. Conclusions

Comparing with nucleus-nucleus collisions, *p-p* collisions are the relatively simpler processes. In these processes, the leading particles are better defined and understood. That is, in each time of *p-p* collisions, there are only two leading particles. They are respectively in the projectile and target fragmentation region.

The charged particles produced in *p-p* collisions are composed of two parts. One is from the hot and dense matter created in collisions, which is supposed to expand and freeze out into measured charged particles according to evolution-dominated hydrodynamics. The other is from the leading particles, which are supposed to have a Gaussian rapidity distribution in their respective formation regions.

Comparing with the experimental measurements performed in *p-p* collisions in the whole available energy region from $\sqrt{s}$ =23.6 to 900 GeV, we can get the conclusions as follows:

(1) The evolution-dominated hydrodynamics is not only applicable to nucleus interactions but also amenable to hadron reactions. This is in accordance with the investigations presented in Ref. [12]. Where, both nucleus-nucleus and *p-p*($\bar{p}$) collisions are assumed to possess the same mechanism of particle production, namely a combination of the constituent quarks in participants together with hydrodynamics. The measurements in nucleus-nucleus reactions are shown to be well reproduced by the measurements in *p-p*($\bar{p}$) interactions. The investigations of Ref. [13] also show that, at high energy, hydrodynamics may be suitable for smaller systems, such as *p-A* and *p-p* collisions.

(2) The centers of the Gaussian rapidity distributions of leading particles increase with energy. However, the widths of distributions are irrelevant to energy.

(3) Though there are only two leading particles in *p-p* collisions, they are essential in describing experimental observations. Only after these two leading particles are taken into account, can the experimental data be matched up properly.

**Figure 1**

The pseudorapidity distributions of the charged particles produced in *p-p* collisions at $\sqrt{s}=$ 23.6, 45.2, 200, 410, 546 and 900 GeV, respectively. The solid dots are the experimental measurements [21-24]. The dashed curves are the results got from the evolution-dominated hydrodynamics of Eq. (9). The dotted curves are the results obtained from the leading particles of Eq. (10). The solid curves are the results achieved from Eq. (13), that is, the sums of dashed and dotted curves.

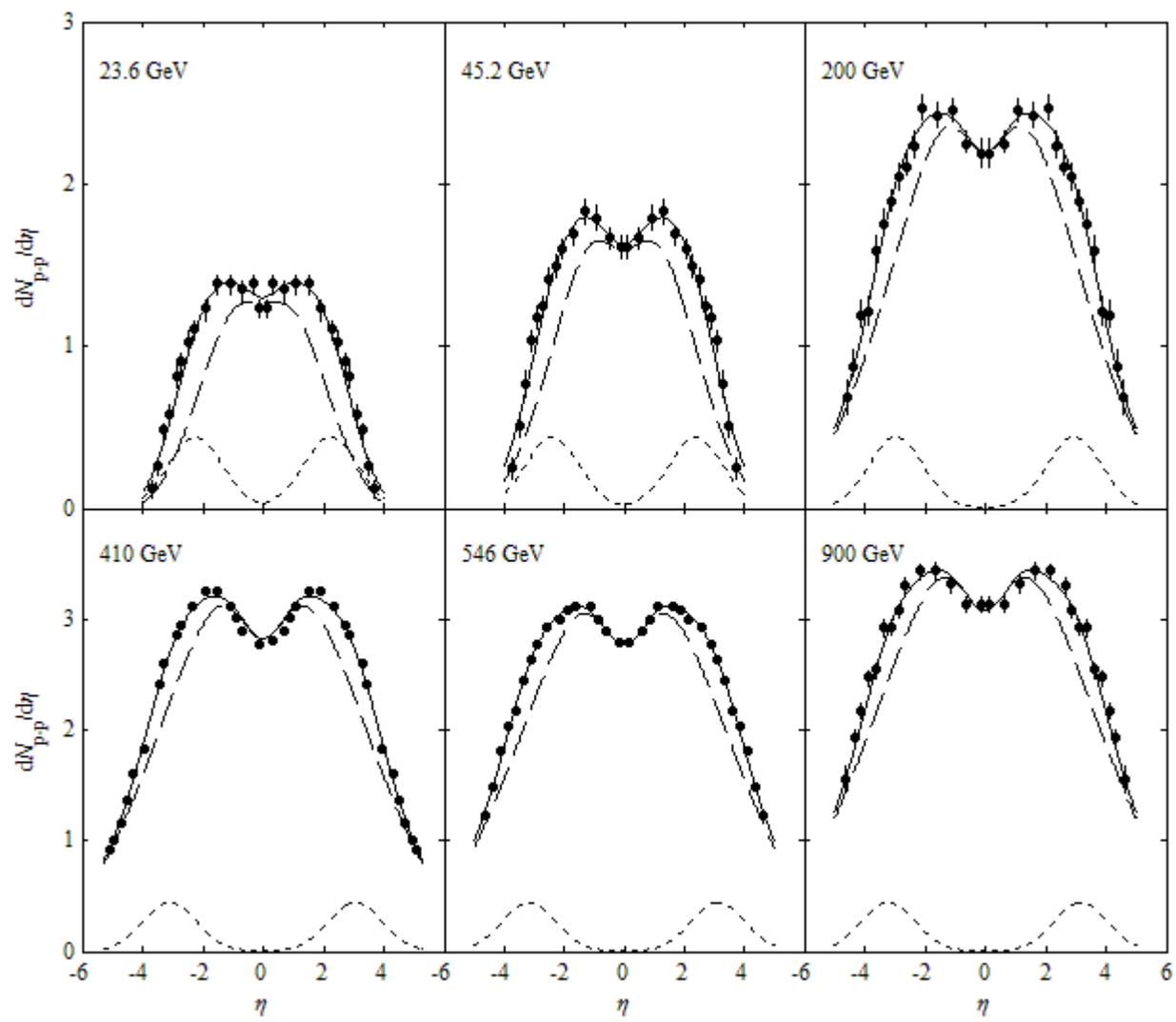

**Figure 1**